\documentclass[conference]{IEEEtran}

\usepackage{graphicx}  
\begin{document}

\title{Using Artificial Neural Networks to Predict Claim
Duration in a Work Injury Compensation Environment}

\author{\authorblockN{Anthony Almudevar}
\authorblockA{Department of Biostatistics and\\Computational Biology\\
University of Rochester\\
Rochester, New York}
}

%


\maketitle
\thispagestyle{plain}
\pagestyle{plain}

\footnote{This work was supported by The Workplace Health, Safety and Compensation
Commission of Newfoundland and Labrador.}

\begin{abstract}
Currently, work injury compensation boards in Canada track injury
information using a standard system of codes (under the National Work
Injury Statistics Program (NWISP)). These codes capture the medical nature and
original cause of the injury in some detail, hence they potentially contain
information which may be used to predict the severity of an injury and the
resulting time loss from work. Claim duration measurements and forecasts are
central to the operation of a work injury compensation program. However,
due to the complexity of the codes traditional statistical modelling
techniques are of limited value.

We will describe an artificial neural network implementation of Cox
proportional hazards regression due to Ripley (1998 thesis) which is
used as the basis for a model for the prediction of claim duration
within a work injury compensation environment. The model accepts as
input the injury codes, as well as basic demographic and workplace
information. The output consists of a claim duration prediction in the form of
a distribution. The input represents information available when a claim
is first filed, and may therefore be used in a claims management setting.
We will describe the model selection procedure, as well as a procedure for
accepting inputs with missing covariates.
\end{abstract}

%
\IEEEpeerreviewmaketitle

\section{Introduction}

The primary purpose of a worker's compensation program is to reimburse the expense associated
with time loss from work due to injury or illness caused in the place of employment.
The duration of a claim is both a measure of the severity of the medical condition as well as the
principle factor driving costs to the program. There is therefore a motivation to develop
statistical models which can predict claim duration based on information available within
a worker's compensation environment.

In this article a model for the prediction of time loss durations is developed, using data
obtained from the administrative database of the Workplace Health, Safety and Compensation
Commission of Newfoundland and Labrador (WHSCC). The prediction is based primarily on a
system of injury codes developed as part of the National Work Injuries Statistics Program
(NWISP) administered by the Association of Workers' Compensation Boards of Canada (AWCBC).
These codes have been adopted by each provincial workers' compensation
board in Canada. They are used to encode a comprehensive description of the cause of the
injury or illness and it's medical nature. A description of the work injury usually
accompanies a compensation claim on a standardized form. The description is then summarized
using the coding system by trained staff, and entered into an administrative database.

In addition to injury codes, information about age, gender, place of residence
and workplace may be used. All this information is available at the time at which a claim
is filed, so that the prediction can be used to aid claims management.
The information is either submitted with the claims form, or can be linked to a claim within
the administrative database.

Because of the complexity of the prediction inputs, Artificial Neural Networks (ANN) present a
natural framework with which to develop a prediction model. Furthermore, such models
can adhere to basic principles of statistical modelling, measuring both the effect of
covariates, and the remaining variability of the modelled response after covariate control
\cite{Cheng1994, Ripley1994, Titterington1999}.

We note, however, that various ANN based models developed for survival time (or risk) prediction,
primarily in the area of cancer prognosis, have reported no significant advantage
when compared to simpler alternative methodologies such as logistic regression, Cox proportional
hazards regression or classification trees \cite{Ripley1998, Groves1999, Lisboa2003, Delen2005},
although improved predictions for ANN models in this application have been reported in
\cite{Jerez2005}.

For the application considered here, the input data consists of 10 categorical covariates,
with class numbers ranging from 2 to 80, resulting in an input vector too complex
for standard regression techniques. We employ a model based on an
implementation of Cox proportional hazards regression
in which the linear prediction term is replaced with an ANN
\cite{Ripley1998b}. The model can be trained using historical claim data
extracted from the administrative database. An important feature is that
censoring, which occurs when a historical claim used for training is not yet closed,
can be naturally accommodated. An important advantage of this type of model is that
a prediction may consist of a complete duration distribution, so that the statistical
range of a prediction can be easily assessed. One feature of this data is that
claim durations associated with similar sets of inputs may vary considerably, so the ability
to capture this variability is crucial.

The structure of the data is discussed in Section II. The proposed model
is introduced in Section III. Model selection and evaluation are discussed
in Sections IV and V. Section VI deals with the problem of partial inputs,
and the article concludes with Section VII.

\section{Structure of Data}

The variables available as predictors are listed in Table \ref{PredictorList}.

\begin{small}
\begin{table}[h]
\caption{Predictor Variables.}
\begin{center}
\begin{tabular}{lll}
\multicolumn{2}{l}{Variable Name} &  \\
Short & Long & Comments \\ \hline
NOI & Nature of Injury & NWISP injury code \\
POB & Part of Body  & NWISP injury code \\
SOI & Source of Injury & NWISP injury code \\
TOA & Type of accident & NWISP injury code \\
AGE & Age of claimant & Categorized by quartile \\
SEX & Sex of claimant &  \\
SIC & Employer Type & Standard Industrial Classification \\
OCC & Occupation  & WHSCC occupation code \\
PAY & Employer size & Total payroll categorized by quartiles \\
CPC & Region & Employee postal code (1st 3 characters) \\
\end{tabular}
\end{center}
\label{PredictorList}
\end{table}
\end{small}

The response variable of interest is short term time loss (or claim duration).
This value represents the number of weeks, starting from the claim open date,
during which compensation was received. A claim is closed when the claimant
returns to work or is moved to a long term benefit program appropriate for severe
or permanent injury. Censoring plays a role in the analysis of the data,
since at the time the data is extracted for analysis there will be claims still
open. The censoring is assumed to be noninformative. The maximum claim open date in the
extract was December 4, 2000, which is the capture date of the data used in this project.

\subsection{Codes}

The main covariates consist of codes used to classify the injury, the place of employment
and the occupation of the claimant. Under NWISP standards injuries are coded
using five distinct codes, which are used to classify {\it part of body} (POB),
{\it nature of injury} (NOI), {\it source of injury} (SOI)
and {\it type of accident} (TOA). The fifth ({\it secondary source of injury})
is not implemented, so we use the remaining four. These codes have a
hierarchical structure. In particular, the NWISP injury codes
consist of five digits, in which greater level of detail is given by lower order digits.
For example, any POB code beginning with '34' signifies an injury
to the fingers. The digit '0' denotes a lack of further detail, so that 34000 is interpreted
as an injury to the finger of fingers for which further detail is unavailable (with respect
to the part of body injured). The code 34001 signifies specifically an injury to the thumb.
Further examples and descriptions are given in Table \ref{CodingExamples}.

Employee type (SIC) is classified using the 1980 Standard Industrial Classification
(SIC) (Statistics Canada) which uses four digits. Occupation type (OCC) is classified using 1980
Canadian Workplace Injury Statistics Program (CWISP) and uses four digits.
These codes have similar hierarchical structure.

The number of distinct codes found in the modelling extract
of each type is given in Table \ref{Consolidation}. In many cases, codes are represented by
a single claim in the extract. An {\it ad hoc} consolidation procedure was
undertaken to combine under-represented categories. Consolidation was done on
the basis of the similarity of code descriptions implicit in the hierarchical
structure. The number of distinct categories after consolidation is given in
Table  \ref{Consolidation}.

\begin{center}
\begin{small}
\begin{table}
\caption{Coding examples}
\begin{tabular}{ll}
\multicolumn{2}{l}{Nature of Injury - Medical condition underlying injury} \\ \hline
 03400 &  Cuts, lacerations \\
 05302 &  Second-degree heat burns, scalds \\
 14530 &  Silicosis \\
 \\
\multicolumn{2}{l}{Part of Body - Location of injury} \\ \hline
 22100 & Esophagus \\
 34000 & Finger(s), fingernail(s), unspecified \\
 34001 & Thumb or thumb and other finger(s) \\
 34002 & Fingers, except thumb \\
\\
\multicolumn{2}{l}{Source of Injury - Object or substance causing injury} \\ \hline
 06400 & Herbicides, unspecified \\
 06410 & Benzoic and phenylacetic acids \\
 06420 & Bipyridyls \\
 11200 & Barrels, kegs, drums \\
 11400 & Boxes, crates, cartons \\
 02100 & Bookcase \\
\\
\multicolumn{2}{l}{Type of Accident - Circumstances of injury} \\ \hline
 02100 & Struck by falling object \\
 11230 & Fall from loading dock \\
 31500 & Contact with overhead powerlines \\
\\
\multicolumn{2}{l}{Standard Industrial Classification - Employer classification} \\ \hline
0617 & Iron Ore and Gypsum Mining \\
4411 & Project Management, Construction \\
\\
\multicolumn{2}{l}{Occupation Codes - Occupation classification} \\ \hline
3115 & Veterinarian \\
8148 & Labourer: Metal Processing \\
\end{tabular}
\label{CodingExamples}
\end{table}
\end{small}
\end{center}

\begin{table}
\begin{small}
\begin{center}
\caption{Coding consolidation}
\begin{tabular}{lcc}
Code & \multicolumn{2}{c}{No. Distinct}  \\
     &  Original & Consolidated  \\  \hline
NOI  & 154 & 61   \\
POB  & 119 & 58   \\
SOI  & 655 & 72  \\
TOA  & 169 & 28   \\
SIC  & 373 & 80 \\
OCC  & 311 & 70  \\
CPC  & 89 & 33  \\
\end{tabular}
\label{Consolidation}
\end{center}
\end{small}
\end{table}

\subsection{Other predictors}

There are two quantitative predictors, age (at claim open date) and employer size
given by total payroll. These are represented by categorical variables AGE and PAY (Table 1).
The values are partitioned into quartiles, then AGE or PAY indicate the quartile into which the
claim belongs. The quartile boundaries for AGE were 30, 38 and 47 years of age. In addition,
the claimants' sex is available as a predictor (SEX).
Within the extract 64.7\% of claimants were male. The employee region is coded using the
first three characters of
the six digit postal code, denoted CPC.

\section{Statistical Modelling}

The data used to predict durations will contain records for which
claims are still open, hence some proportion of the responses used to build the
model will be censored, in the sense that the claim duration extracted
from the data base represents only a lower bound for the eventual duration,
unless the claim is indicated as being closed. In Figure \ref{LongTermTrendGraph} the proportion
of claims still open as of the capture date of December 4, 2000 is shown as a
function of open date. Approximately 30\% of recently opened claims remain open,
and approximately 5\% of claims opened 5 years previously remain open, hence techniques
from survival analysis which account for censoring will be indispensable.

A random lifetime (a positive valued random variable) has density $f(t)$ and
cumulative density function $F(t)$. The \emph{hazard rate} is defined to be
$h(t) = f(t)(1-F(t))^{-1}$, and can be interpreted as the instantaneous rate
of death at time $t$ conditional on survival up to that time. It is a complete
representation of $f(t)$ (for reference, an exponential density, which models
a 'memoryless survival time', has a constant hazard rate, so
that the death rate is independent of current age).

Suppose we wish to model dependence of the hazard rate function on a covariate
vector of fixed structure $x$ (a $p \times 1$ real valued column vector).
Let $\eta(x)$ be a real valued prediction term. The Cox proportional hazards model
\cite{Cox1972} takes the form
\begin{eqnarray}
h_x(t) & = & h_o(t)e^{\eta(x)} \label{CoxModel}
\end{eqnarray}
where $h_o(t)$ is the \emph{baseline hazard rate}.
The implied \emph{proportional hazards assumption} is that all hazard rates
are identical in shape to $h_o$ up to a multiplicative constant. When this
assumption holds, this model is particularly tractable, since the likelihood
conditioned on the times of death does not depend on $h_o$. Then $\eta$
can be estimated with the available data. If needed, $h_o$ can be recovered from
the fitted model. In the most common implementation, $\beta$ is a $p \times 1$
column vector, and  $\eta = \beta^T x$ is the linear prediction term. The partial
likelihood is then maximized with respect to $\beta$. However, the structure
of the likelihood does not depend on $\eta$, so more flexible prediction
terms may be used. For example, in the \emph{generalized additive model} form
proposed in \cite{Hastie1990} $\eta$ assumes nonparametric functional forms.
Here, we adopt the form proposed in \cite{Ripley1998b}, in which $\eta$ is the
output of an ANN.

\subsection{Long term duration trends}

\begin{figure}
\center \scalebox{0.5}{\includegraphics{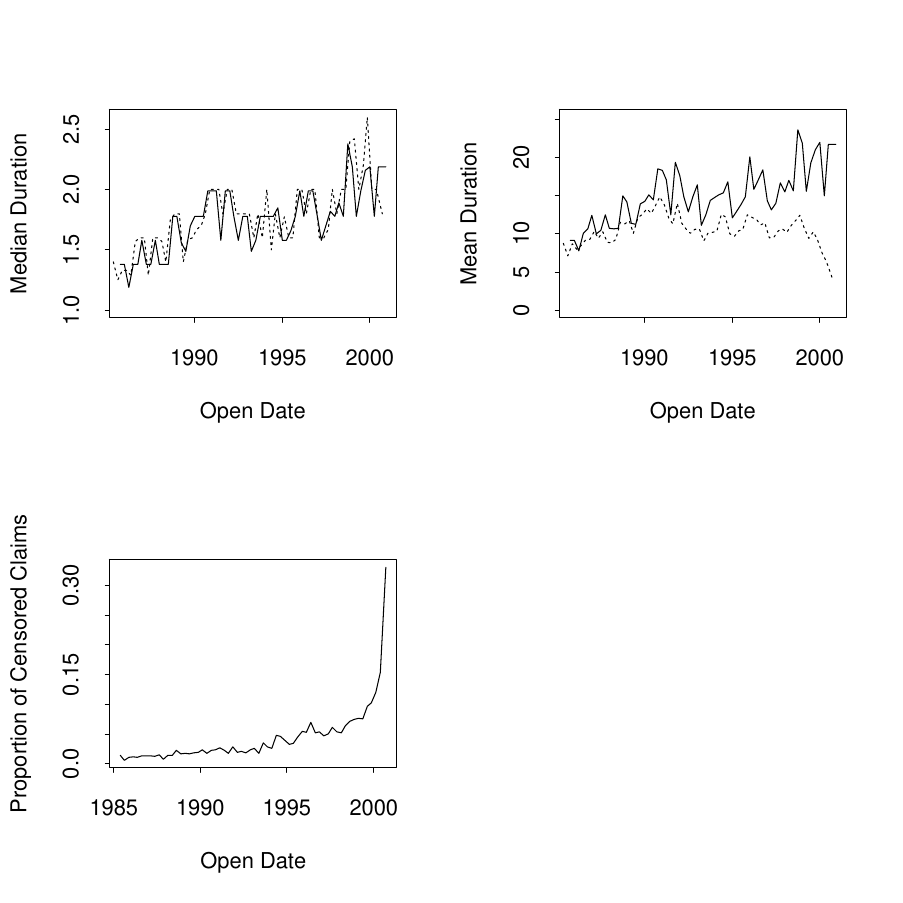}}
\caption{Long term trend in claim duration. 1) and 2) Mean and
median trends using piecewise linear Cox regression (solid) and
quarterly summaries (dashed). 3) Quarterly summaries of censoring
rates.}
\label{LongTermTrendGraph}
\end{figure}

The graphs in Figure \ref{LongTermTrendGraph} were obtained by fitting a piecewise linear
Cox regression model of mean and median claim duration against claim open date. Continuous knots were imposed at
the end of yearly quarters. For comparison, quarterly summaries of mean and median duration
are shown (not accounting for censoring). The
quarterly censor rates are also given. There is a clear upward trend in claim duration, evident
when methods which account for censoring are employed. The quarterly mean summaries do not account
for censoring, and so give an inaccurate depiction of falling claim durations. The median summaries,
not as sensitive to censorship as the mean summaries, roughly conform to censoring-specific estimates.
We also note that the current NWISP injury codes replaced an earlier standard prior to 1998.
Thus, in order to use a data set which is time-homogeneous in duration distribution and coding
protocols the prediction model used data with an open date of January 1, 1998 or later.
Finally, records with durations of 0 were removed (such claims may receive compensation for the cost
of medical treatment but experience no time loss). This resulted in a total of 17,026 claims
available for modelling (the modelling extract).

For the purposes of model selection the 17,026 were randomly divided into
a training sample (10,000 claims) and a test sample (7,026 claims). In general, models
were fit using the training sample, then were evaluated using the test sample.

\subsection{Proportional hazards assumption}

\begin{figure}
\center \scalebox{0.45}{\includegraphics{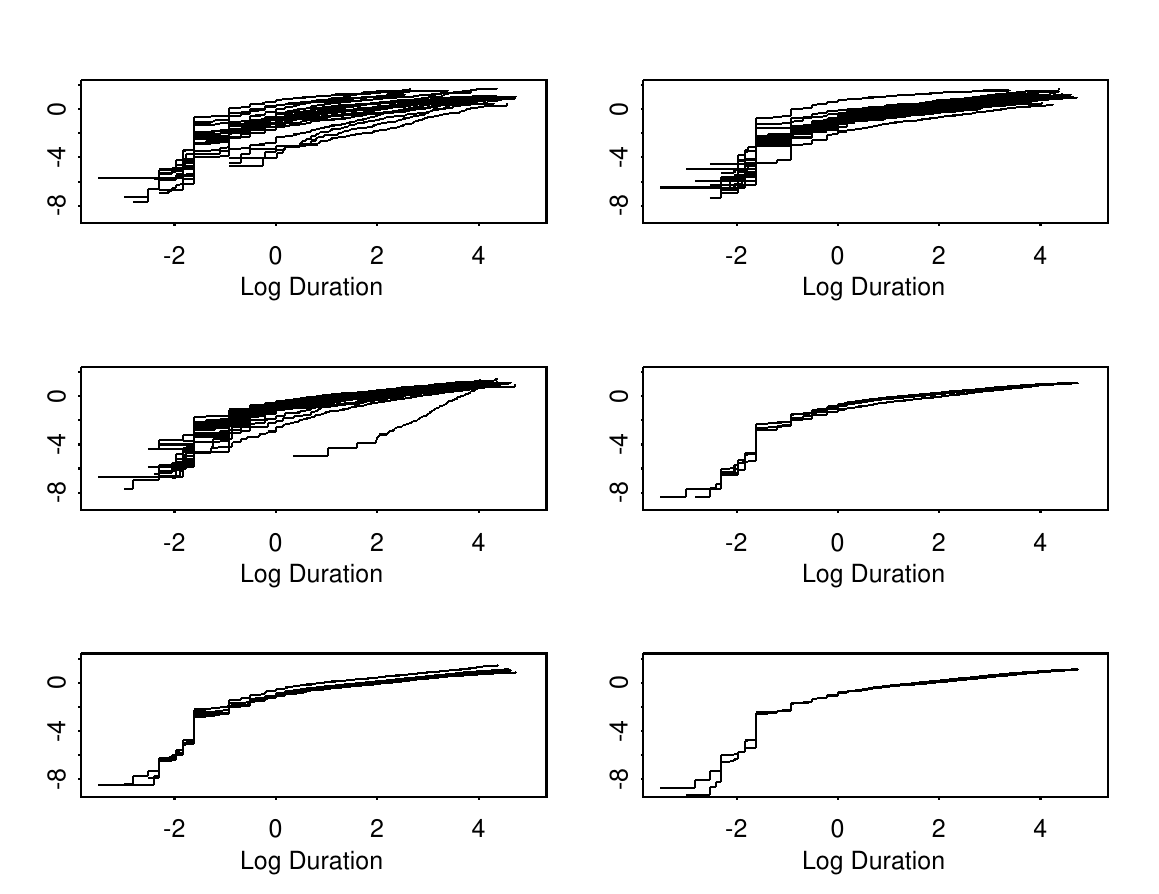}}
\caption{Log cumulative hazard functions partitioned by NOI, POB,
OCC, PAY, AGE and SEX (by column, then by row).}
\label{PHassumption}
\end{figure}

To validate the proportional hazards assumption, the complete data
set was partitioned in turn by the six codes indicated in Figure
\ref{PHassumption}. For each subset, the cumulative hazard
function $H(t) = \int_0^t h(u)du$ was estimated using a
Kaplan-Meier estimate of the survivor function $S(t) =
\exp(-H(t))$ (equivalent to the population proportion which
survive past time $t$). Under the model of equation \ref{CoxModel}, the log
cumulative hazard functions will differ by a constant. This effect
is evident in Figure \ref{PHassumption}.

\subsection{Main effects model}

\begin{figure}
\center \scalebox{0.45}{\includegraphics{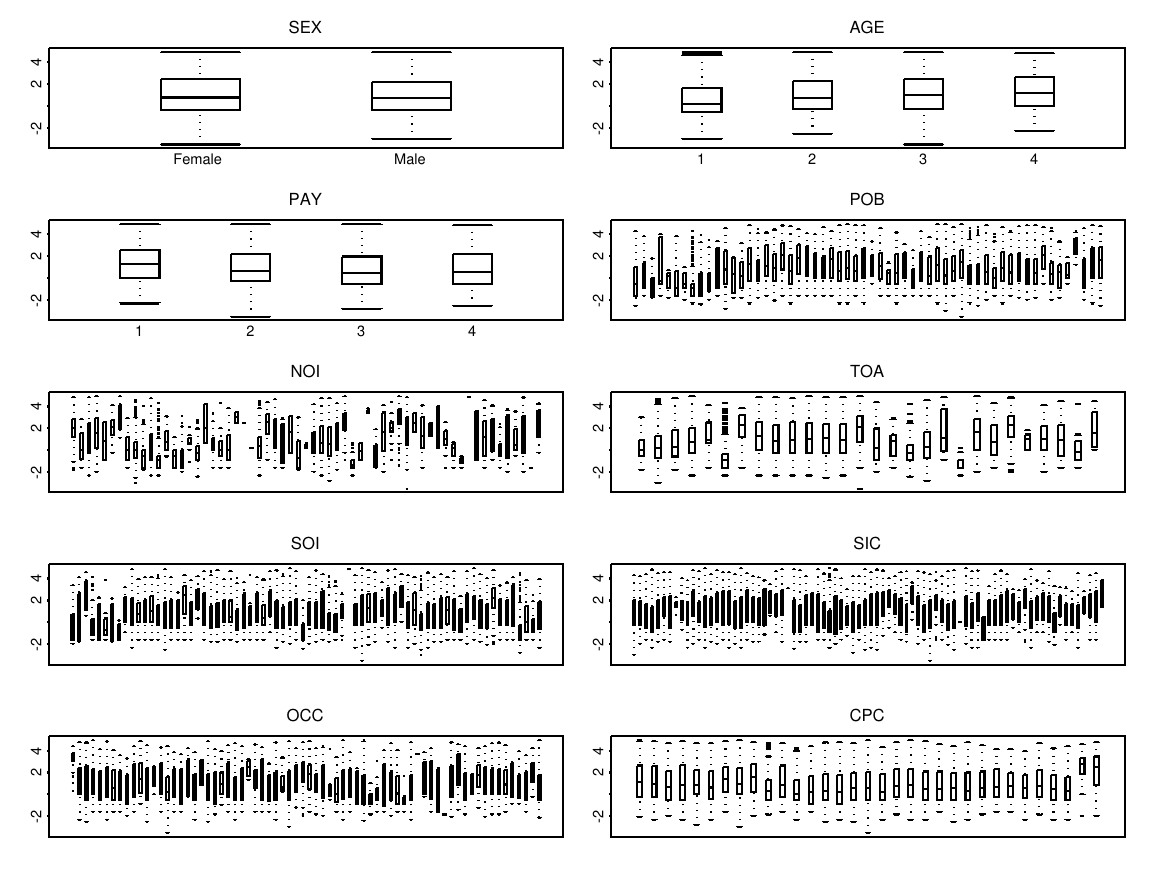}}
\caption{Multiple boxplots of log duration given for each predictor.}
\label{MainEffectsBoxplots}
\end{figure}

Multiple boxplots of log durations are given in Figure \ref{MainEffectsBoxplots},
giving an indication of the degree to which the predictors affect duration.
In order to compare the proposed ANN model to a simpler alternative, a
\emph{main effects} Cox proportional hazards model was developed, treating each of
the 10 predictors as a factor. This gives a model structure similar to multiple treatment
analysis of variance (ANOVA), using no treatment interactions. The model was built by
adding predictors successively. Using likelihood ratio tests for the nested sequence,
all predictors were associated with significant increases in likelihood.

The assumption behind a main effects model is that the covariates influence
duration independently of each other. For example, if durations
for females are higher than for males, then such a model will always
predict higher duration for a female when all other covariates are otherwise
equal. However, we find that the sex ordering of durations depends on the
covariates. To illustrate this, the entire data set was partitioned by NOI
code. For each subset in which each sex is represented at least 10 times, a
log-rank test \cite{Harrington1982} was used to detect significant differences in duration
by sex. For each significant result (P-value $\leq 0.05$),  the sex with the
highest durations was determined. Of the 25 codes satisfying the size
cutoff, the significant differences are listed in Table \ref{NOIxSEX}. A similar
analysis was carried out for POB, with 38 codes satisfying the size cutoff
(Table \ref{POBxSEX}). We conclude that sex ordering interacts with the injury
codes.

The presence of interactions means that a main effects model would fail to
capture potentially important structure. Interactions can, in principle,
be added to the model, but the large number of potential interactions, given
as products of the number of categories in Table \ref{Consolidation}, would make this
process far too complex. As an alternative, an ANN would be suitable
for modelling these interactions using potentially far fewer parameters.

\begin{table}
\begin{small}
\begin{center}
\caption{Interactions between Nature of Injury and Sex (out of 25 codes)}
\begin{tabular}{llr}
  & Sex with & \\
NOI & longer durations &  P value \\ \hline
Dislocations & Female & .0025 \\
Sprains, strains, tears & Female &  $<$ .0001 \\
Rheumatism, except the back, unsure & Female  & $<$ .0001 \\
Avulsions & Male & $<$ .0001 \\
Burns, n.e.c. & Male & .0081
\end{tabular}
\label{NOIxSEX}
\end{center}
\end{small}
\end{table}

\begin{table}
\begin{small}
\begin{center}
\caption{Interactions between Part of Body and Sex (out of 38 codes)}
\begin{tabular}{llr}
  & Sex with & \\
POB & longer durations &  P value \\ \hline
Chest, except int. loc. of diseases ... & Female & .0373 \\
Wrist(s) & Female & .0003 \\
Knee(s) &  Female & .0106 \\
Groin  & Male  & .0283 \\
Finger(s), fingernail(s) & Male & $<$ .0001 \\
Multiple body parts & Male  & .0499
\end{tabular}
\label{POBxSEX}
\end{center}
\end{small}
\end{table}


\subsection{ANN model}

We will use an implementation of the Cox proportional hazards model
due to \cite{Ripley1998b} in which the predictor term $\eta$ of (\ref{CoxModel})
is calculated using an ANN, using the inputs defined in Table \ref{PredictorList}.
The network architecture is based on a multilayer perceptron, structurally identical
to a directed graph $(V,E)$ with $n_t$ nodes $V$ labelled $i = 1,\ldots,n_t$. Edge $(i,j) \in E$ exists
if node $i$ connects to node $j$. Node $i$ is labelled by a real number $x_i$, and edge $(i,j)$
is labelled with real-valued weight $w_{ij}$.

Nodes are arranged in three disjoint layers, an input
layer of $n_i$ nodes, a hidden layer of $n_h$ nodes, and an output layer
consisting of a single node. The network implemented here
is \emph{fully connected}, that is, each input node is
connected to each hidden node, and each hidden node is connected to the output node.
Optionally, each input node is connected to the output node
({\it skip layer} connections), which are implemented here. No other type of connection is permitted.
Denote the input and hidden nodes $I$ and $H$, and input-to-hidden, hidden-to-output and
input-to-output edges $IH$, $HO$ and $IO$ respectively.

Labels for the hidden layer nodes are given by the formula
\begin{eqnarray*}
x_{j^\prime} & = & \phi\left(  \sum_{(i,j) \in IH : j = j^\prime} w_{ij^\prime} x_i \right), \,\,\, j^\prime \in H,
\end{eqnarray*}
where $\phi(u) = \exp(u)/(1+\exp(u))$. An input vector $x$ is an ordered list of
$n_i$ real numbers to be assigned as labels for the input nodes. In \cite{Ripley1998b}
this generates a relationship between the input vector and the output node label:
\begin{eqnarray}
\eta(x) & = & \sum_{(i,j) \in IO} w_{ij} x_i \nonumber \\
       & &  + \sum_{(i,j) \in HO} w_{ij} \phi\left( \sum_{(i,j) \in IH} w_{ij} x_i \right). \label{net}
\end{eqnarray}
One input node (the {\it bias node}, labelled 0) is added which accepts
a constant input of 1.  Note that the transformation $\phi$ is not applied to the output node.

The model is constructed by allowing the term $\eta(x)$ of (\ref{net}) to be
incorporated into the model of equation (\ref{CoxModel}), in place of the more commonly used
linear prediction term. In the model proposed here, each input variable is categorical
(AGE and PAY are given as quartiles, and so have four classes, while SEX has two classes). For
each categorical variable, one input node is assigned to each class (category value). A set
of categorical variables is transformed into an input vector by setting the input node
assigned to a categorical variable class equal to one, and the remaining nodes equal to zero.
The number of nodes in the input layer is therefore determined by the structure of the input vector $x$.
On the other hand, the number of nodes in the hidden layer can be varied in the context of a model selection
procedure. For a given number of hidden nodes, given training data consisting of input
vector/survival time pairs, the weights are determined by minimizing
\begin{eqnarray}
\lefteqn{  H(W|X_l,Y_l,\lambda,\lambda_b,n_h)  }  & & \nonumber \\
 & &  =  - L(W,X_l,Y_l)
    + \lambda \sum_{i \geq 1} w^2_{ij}
    + \lambda_b \sum_{i = 0} w^2_{ij} \label{objective}
\end{eqnarray}
with respect to $W$ where $W$ represents all weights $\{w_{ij} : (i,j) \in E\}$,
$X_l,Y_l$ represents all predictors and (censored) responses in the training data,
$\lambda$ is the {\it decay} parameter and $\lambda_b$ is the {\it bias decay} parameter.
The parameters  $\lambda$ and $\lambda_b$ function as smoothing parameters, with a separate
value applied to weights connected to the bias node. We follow the practice of setting
$\lambda_b$ to be a small fraction of $\lambda$, here 1/25 \cite{Ripley1998b}.
The quantity $L(W,X_l,Y_l)$ is the Cox partial likelihood defined in \cite{Cox1972},
which follows from equation (\ref{CoxModel}) and (\ref{net}).

\section{Model Selection}

The model defined by (\ref{CoxModel}), (\ref{net}) and (\ref{objective})
can be varied in three ways. First, we consider reduced and full models:
\begin{itemize}
\item[(R)] AGE, SEX, POB
\item[(F)] AGE, SEX, POB, NOI, TOA, SOI, SIC, OCC, CPC, PAY.
\end{itemize}
The model may be further varied by altering the decay parameter $\lambda$
and the number of hidden nodes $n_h$.

\begin{figure}
\center \scalebox{0.45}{\includegraphics{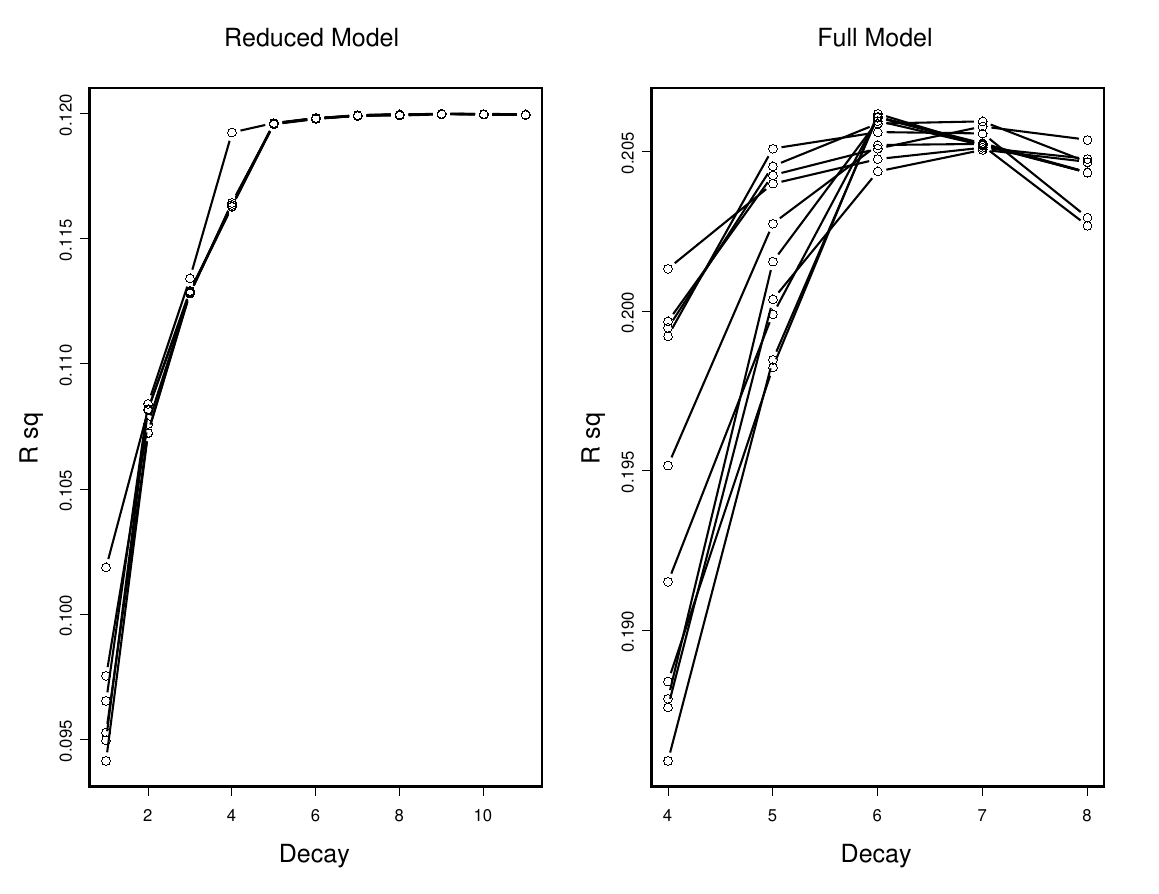}}
\caption{Generalized $R^2$ by decay parameter, for reduced and full models.
Separate lines corresponding to number of hidden nodes are superimposed.}
\label{ModelSelection}
\end{figure}

To score a model, a fit was calculated using the 10,000 training records
to minimize (\ref{objective}) for fixed $n_h, \lambda$ and model size (R) or (F).
The resulting weights and baseline hazard function were used to implement
(\ref{CoxModel}) and (\ref{net}). The prediction term $\eta$ of (\ref{net}) was then
calculated for each of the 7,026 test records. A new univariate Cox proportional
hazards models was calculated for the test records, using $\eta$ as a linear predictor.
The fit of the model was evaluated using the generalized $R^2$ coefficient
\begin{eqnarray*}
R^2 & = & 1-\exp(- 2 (\log(\ell_{full}) - \log(\ell_{null})) / n)
\end{eqnarray*}
where $\ell_{full}$ and $\ell_{null}$ are the full and null likelihoods of the univariate
fit, and $n$ is the number of records. This quantity has a similar interpretation as
the coefficient of determination in univariate linear regression \cite{Nagelkerke1991}.
Note that this quantity is based on a model of one degree of freedom, thus controlling
for overfitting. Figure \ref{ModelSelection} displays $R^2$ for the full and reduced models
over a range of $\lambda$ and $n_h$. On this basis, a full model was selected with
$n_h = 12$, $\lambda = 6$, with $R^2 = 20.6$. This model employed 426 nodes.
Up to $n_h = 14$ hidden nodes were evaluated. For comparison, the same procedure
was carried out using the main effects model described
in Section III-C, using the variable subset of the full model (F),
which resulted in $R^2 = 0.15$.

\section{Model Evaluation}

\begin{figure}[h]
\center \scalebox{0.45}{\includegraphics{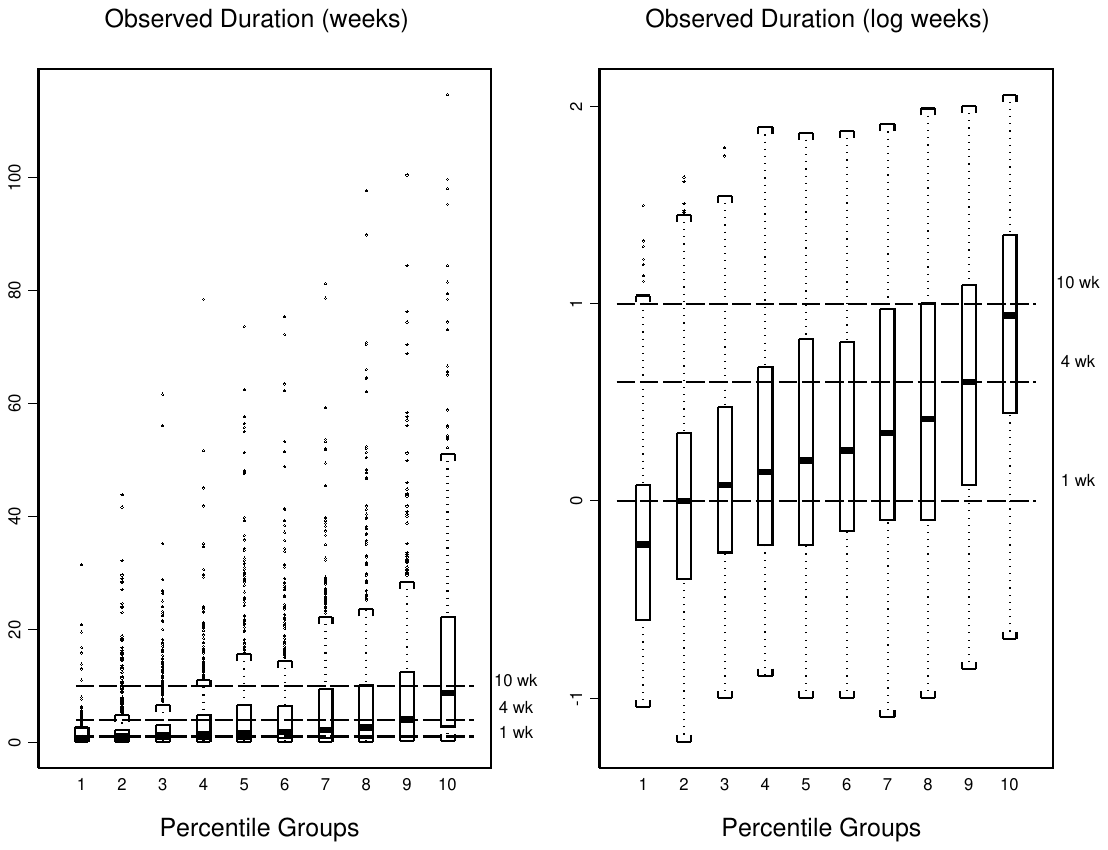}}
\caption{Actual durations (for closed claims) partitioned into deciles
on the basis of prediction terms calculated from selected model. Plots
are given in original and logarithm scale, for clarity. The location
of 1, 4 and 10 weeks is indicated.}
\label{BoxplotDeciles}
\end{figure}

To evaluate the model, the prediction term $\eta$ was calculated
for each test record using the model selected in Section IV. Note that these
prediction terms will possess the same ordering as the duration that would be
predicted by substituting $\eta$ into (\ref{CoxModel}). The prediction terms
were partitioned into deciles, and  actual test durations
summarized by boxplots for each decile-partition in Figure \ref{BoxplotDeciles}. A clear
relationship between predicted and actual duration is indicated.
Conditional probabilities for actual quintiles given
predicted quintiles are given in Table \ref{QuintileTable}, which also
indicate good predictive ability.

\begin{table}
\begin{small}
\begin{center}
\caption{Proportion of actual closed duration quintiles by predicted quintiles.}
\begin{tabular}{cccccc}
Prediction  & \multicolumn{5}{c}{Actual Duration Quintile}  \\
Quintile & 1 &   2 &  3 & 4 & 5 \\ \hline
1 &  0.38 &  0.27 &  0.20 &  0.11 &  0.04 \\
2 &  0.25 &  0.22 &  0.23 &  0.19 &  0.12 \\
3 &  0.19 &  0.21 &  0.20 &  0.23 &  0.17 \\
4 &  0.16 &  0.16 &  0.19 &  0.22 &  0.27 \\
5 &  0.09 &  0.12 &  0.13 &  0.26 &  0.40
\end{tabular}
\label{QuintileTable}
\end{center}
\end{small}
\end{table}

\begin{figure}[h]
\center \scalebox{0.45}{\includegraphics{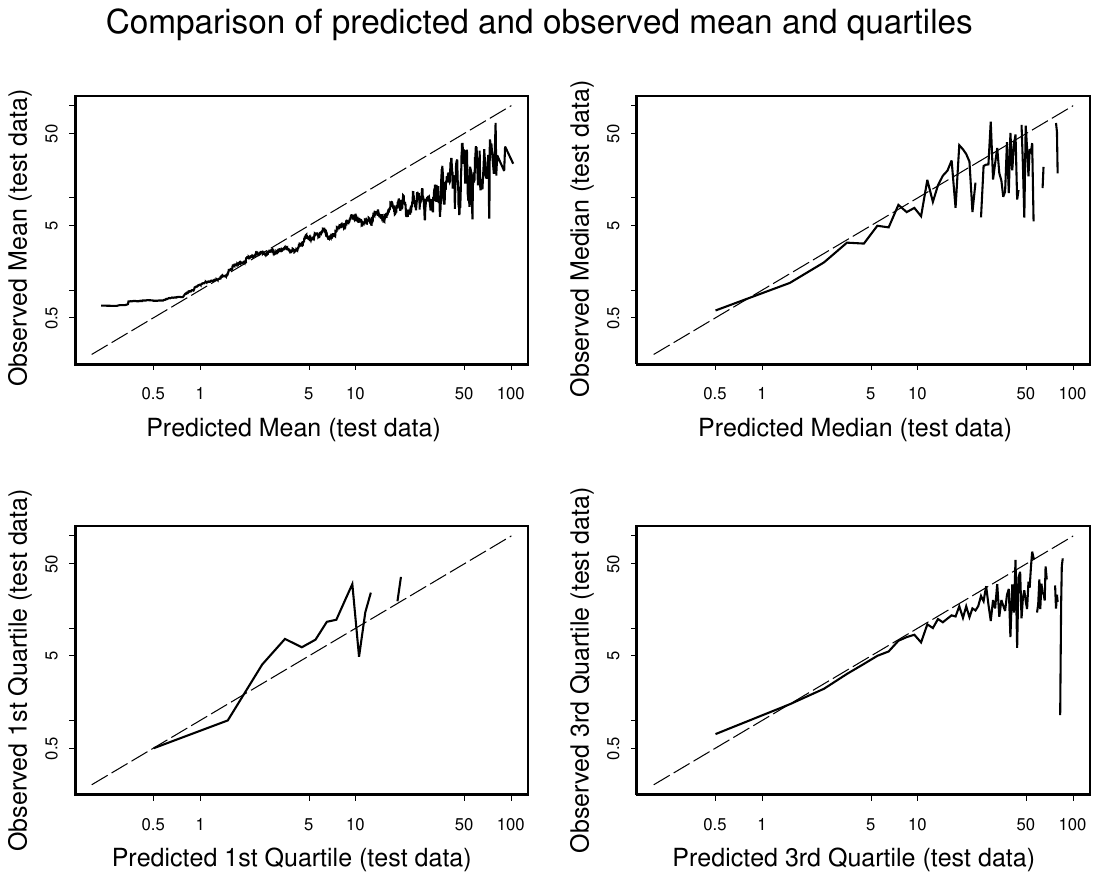}}
\caption{Estimated relationship between actual and predicted duration
means and quartiles by individual subject.}
\label{DistributionAnalysis}
\end{figure}

Because (\ref{CoxModel}) permits the calculation of a predicted distribution
directly from $\eta$, a subject-level comparison of predicted and actual duration distributions
can be made. For a given $\eta$ predicted distributional summaries,  such
as means and quartiles obtained from (\ref{CoxModel}), can be compared to analogous summaries
from the actual durations conditioned on the predicted quantities.
This can be done by constructing a moving window of one week duration centered at a given
predicted value. The actual summary value is calculated within the window.
The results are shown in Figure \ref{DistributionAnalysis}. Some bias is
indicated for the mean, but the quartiles are accurately estimated, although
for each summary, the estimates become unstable for larger durations.

\section{Predictions Based on Partial Inputs}

It may be impractical to require the complete 10 predictor inputs of Table I
in order to make a prediction. We therefore consider two procedures for making
predictions based on partial inputs.

{\bf Method A}: When a partial input is made, the application takes the average value of
prediction term $\eta$ among all training records matching the partial input.
This average becomes the prediction term.

{\bf Method B}: When a partial input is made, the application takes the average survival
curve among all training records matching the partial input. This becomes the output survival curve.

\begin{figure}
\center \scalebox{0.45}{\includegraphics{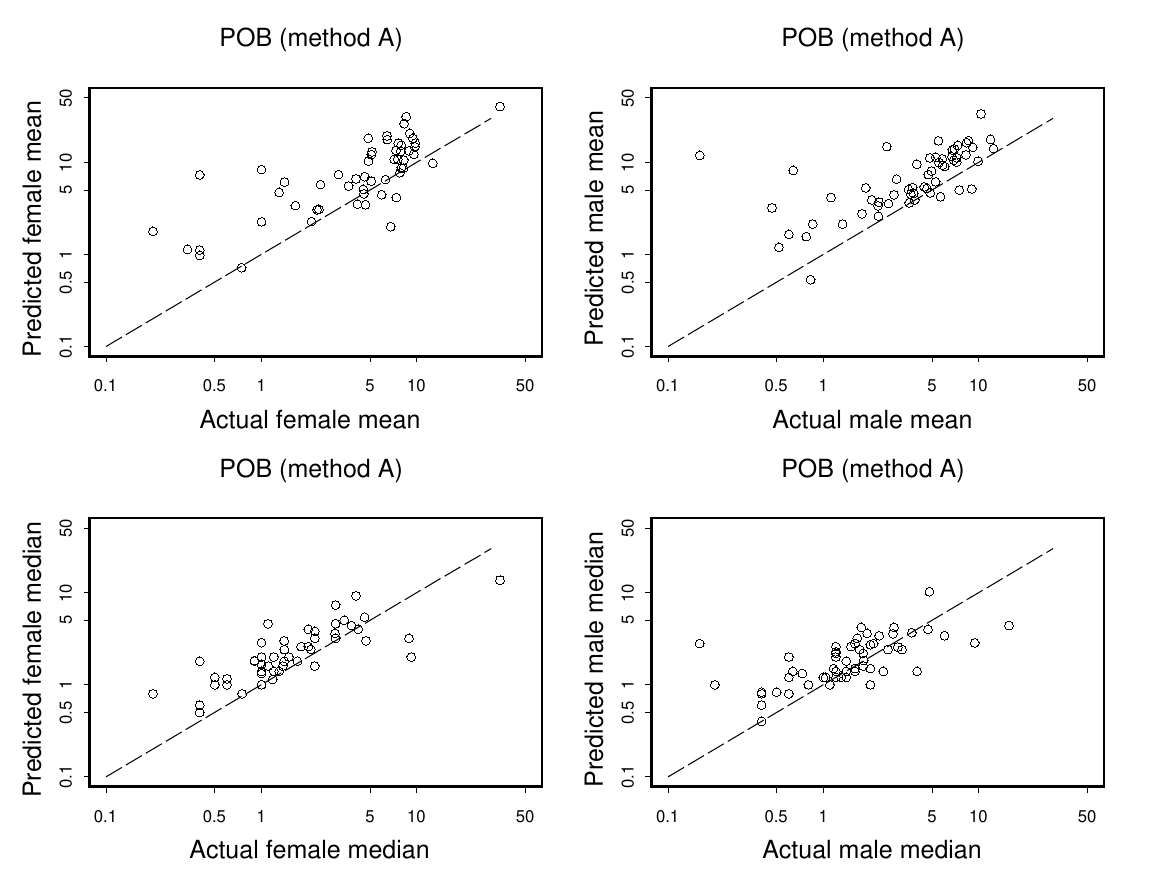}}
\caption{Predicted mean and median duration against actual mean and median
duration partitioned by POB, for each sex. Predictions are calculated
using Method A (averaging over prediction terms).}
\label{MethodA}
\end{figure}

\begin{figure}
\center \scalebox{0.45}{\includegraphics{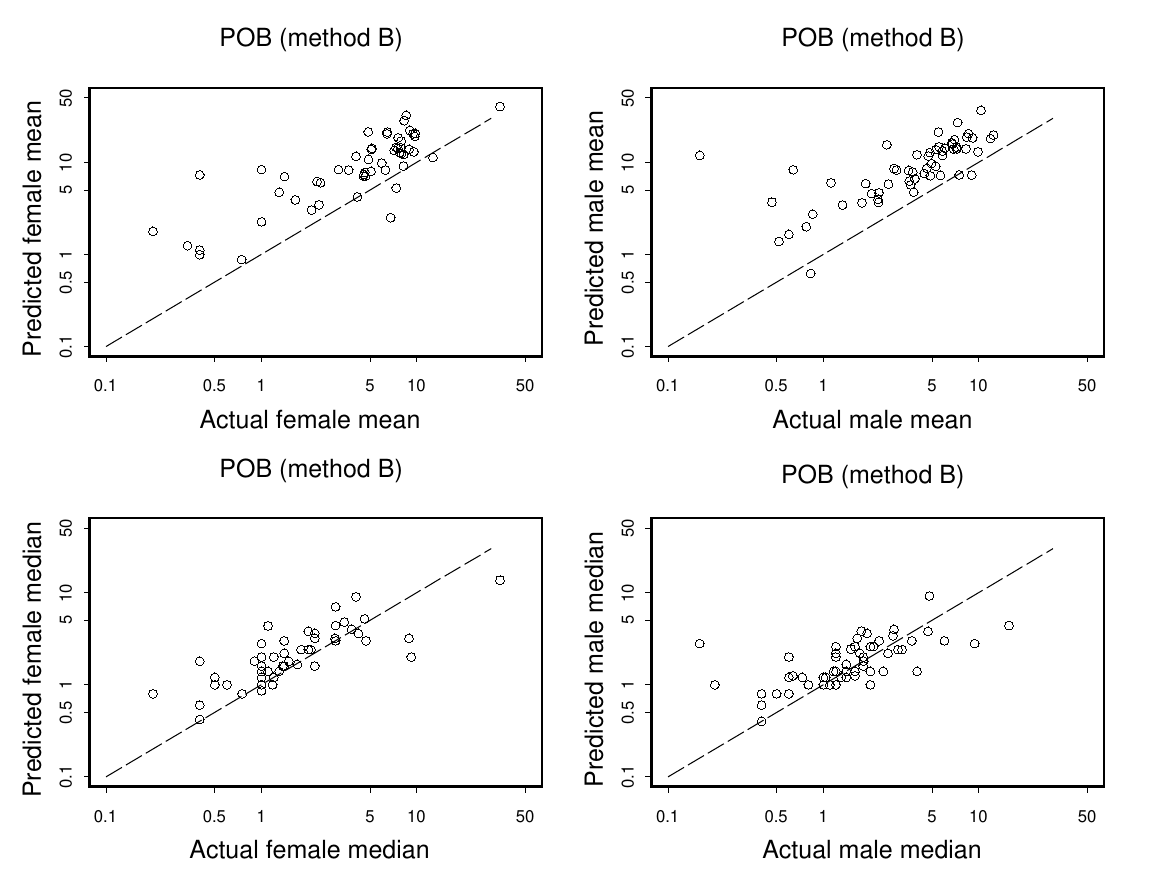}}
\caption{Predicted mean and median duration against actual mean and median
duration partitioned by POB, for each sex. Predictions are calculated
using Method B (averaging over survival curves).}
\label{MethodB}
\end{figure}

To compare these methods, predictions were calculated using each combination
of POB code and SEX as the partial input. Means and medians were calculated and compared
to the actuals values obtained from the data. Here we use the training data.
Figures \ref{MethodA} and \ref{MethodB} display the results. A close correspondence
between the predicted and actual summaries is evident, although there is a tendency
for the prediction calculated using Method B to overestimate the mean. Additionally, Method A
is simpler to implement, so we adopt this procedure.

\begin{figure}[h]
\center \scalebox{0.45}{\includegraphics{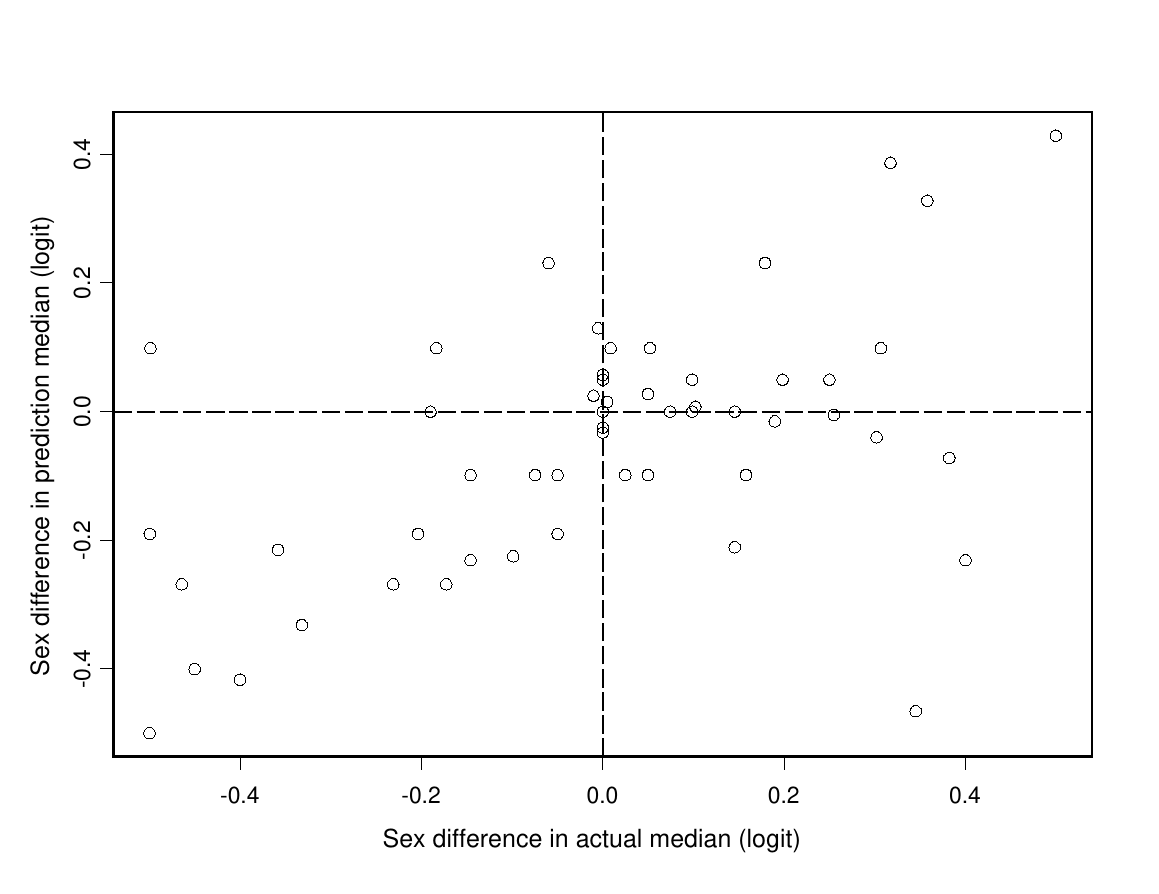}}
\caption{Sex difference in median duration partitioned by POB, predicted
versus actual. Durations are transformed using a centered logistic
function.}
\label{IntA}
\end{figure}

Finally, we examine the ability of the model to capture interactions.
For each POB code, the actual sex difference in median duration
(male minus female) is compared to the the predicted difference, using the
training data. A scatter plot is given in Figure \ref{IntA}, indicating a
tendency for the predicted sex difference to capture the actual sex difference.
Note that if a main effects model were to be used, the predicted
sex difference would not vary with the other predictor values.
In 30 of 45 codes examined (small samples and zero differences
were not included in the analysis) the predicted difference is of the same
sign as the actual. Kendall's correlation coefficient (based on an excess of
concordant pairs) was calculated for the scatter plot and found to be
significantly different from 0 (P = 0.0003).

\section{Conclusion}

An ANN model was developed for the prediction of claim duration
in a work injury compensation environment using a detailed coded
description of the injury, as well as demographic covariates, as input.
The network is imbedded in a Cox proportional hazard model,
so that censoring of training data may be accounted for,
and a complete duration distribution may be output. Model validation
suggests that the predictions are able to accurately model
both the effect of the input, as well as the inherent variability
of the durations. A procedure for dealing with partial inputs was proposed,
and found to accurately predict durations implied by the input.
The advantage of the ANN approach was demonstrated by the ability
of the model to efficiently capture predictor interactions.
Overall, the model provides an example demonstrating the efficacy of
ANN based modelling in applications for which the complex structure of
the predictor set effectively precludes the use of standard statistical
modelling tools.

\end{document}